\documentclass[prl,twocolumn,floatfix,amsfonts]{revtex4}
\usepackage{graphicx,graphics,color,epsfig}
\usepackage{bm}
\usepackage{amsmath}
\usepackage{amssymb}
\newcommand{\bS}{{\bf S}}

\newcommand{\bk}{{\bf k}}

\newcommand{\br}{{\bf r}}

\begin{document}
\title{Comment on ``Coexistence of Superconductivity and Ferromagnetism in Ferromagnetic 
Metals~\cite{littlewood}''.}
\author{Yogesh N. Joglekar$^1$ and Allan H. MacDonald$^2$}
\affiliation{$^1$ Theoretical Division, Los Alamos National
Laboratory, Los Alamos, New Mexico 87544, \\
$^2$ Department of Physics, University of Texas at Austin, Austin, Texas 78712.}
\date{\today}

\begin{abstract}
We argue that a single-band itinerant electron model with short-range interactions, proposed by 
Karchev {\it et al.}~\cite{littlewood}, cannot describe the coexistence of superconducting and 
ferromagnetic order. 
\end{abstract}
\maketitle


In a recent Letter~\cite{littlewood} Karchev {\it et al.} proposed a model for the coexistence of 
$s$-wave superconductivity and ferromagnetism, in which both orders arise from itinerant electrons. 
It has been investigated further in a recent preprint~\cite{bedell}. The results presented 
in~\cite{littlewood} are based on a model with a {\it local} electron-electron interaction of the form 
$-J{\bS}_{\br}\cdot{\bS}_{\br}/2-gn_{\br\uparrow}n_{\br\downarrow}$. In this comment, we point out 
that the Hubbard-Stratonovich (HS) transformation used in~\cite{littlewood} predicts an ordered state 
even in the case of {\it non-interacting} electrons. Although the HS approach is attractive for the 
physical transparency it brings to the study of quantum fluctuations in ordered states, it tends not 
describe the microscopic competition between different possible types of order well. The evidently 
incorrect inference mentioned above, and the well-known~\cite{negele} property that HS transformation 
can be used to derive Hartree or Fock but not Hartree-Fock mean-field equations, are examples of this 
difficulty. In this comment we show explicitly that the model considered in~\cite{littlewood}, when 
treated by a Hartree-Fock mean-field theory, leads to a physically sensible phase-diagram that does 
not support simultaneous ferromagnetism and $s$-wave superconductivity~\cite{caveat}. 

Using the identity ${\bS}_{\br}\cdot{\bS}_{\br}=3(n_{\br\uparrow}+n_{\br\downarrow})/4-
3n_{\br\uparrow}n_{\br\downarrow}/2$, the local interaction can be written as 
$-\tilde{J}\lambda{\bS}_{\br}\cdot{\bS}_{\br}/2-\tilde{g}(1-\lambda)n_{\br\uparrow}n_{\br\downarrow}$ 
for arbitrary $\lambda$, where $\tilde{J}\equiv(J-4g/3)=-4\tilde{g}/3$. In~\cite{littlewood} ordered 
states can occur even for the non-interacting case, $\tilde{J}=0=\tilde{g}$, indicating breakdown of 
the HS mean-field theory. To cast the subsequent discussion in a transparent Hartree-Fock language, 
we perform a particle-hole transformation on the down-spin, 
$c_\downarrow(\br)\rightarrow d^{\dagger}_\downarrow(\br)$. The Hamiltonian expressed in terms of 
$d$-fermions is given by 
\begin{eqnarray}
\label{eq: one}
H & = & \sum_{\sigma\sigma'\bk}d^{\dagger}_{\bk\sigma}\left[\xi_\bk\tau^{z}- \tilde{g}(1-\lambda)
\mathbf{1}\right]_{\sigma\sigma'} d_{\bk\sigma'}\nonumber \\
& + & \int d\br\, \left[\frac{\tilde{J}}{2}\,\lambda{\cal S}_{\br}\cdot{\cal S}_{\br}+
\frac{\tilde{g}}{2}\,(1-\lambda)(n^{d}_{\br\uparrow}+n^{d}_{\br\downarrow})^2 \right]
\end{eqnarray} 
where $\xi_\bk=\epsilon_\bk-\mu-g/2$ is energy measured from a shifted chemical potential, 
$n^{d}_{\br\sigma} ({\cal S}_{\br})$ is the number (spin) operator for $d$-fermions at position $\br$, 
and we have used the identities ${\bS}_{\br}\cdot{\bS}_{\br}= -{\cal S}_{\br}\cdot{\cal S}_{\br}$ and 
$2n_{\uparrow\br}n_{\downarrow\br}=(3n^{d}_{\uparrow\br}+n^{d}_{\downarrow\br})-
(n^{d}_{\uparrow\br}+n^{d}_{\downarrow\br})^2$ to derive Eq.(\ref{eq: one}). In this language 
$s$-wave superconductivity corresponds to nonzero $\hat{x}$-$\hat{y}$ 
spin-polarization for the $d$-fermions. The local interaction above is the sum of density 
($g_n n_d^2/2$) and isotropic spin-dependent ($g_s{\cal S}\cdot{\cal S}/2$) contributions which give 
rise to Hartree mean-fields $g_n n_d\mathbf{1}$ and $g_s{\vec\tau} \cdot {\vec m}/4$, and exchange 
mean fields $-g_n(n_d\mathbf{1}+{\vec \tau} \cdot {\vec m})/2$ and 
$-g_s(3n_d\mathbf{1}-{\vec \tau}\cdot{\vec m})/8$ respectively. Here 
$n_d (\vec{m})=\int_{\bk}\langle d^{\dagger}_{\bk}\mathbf{1} (\vec{\tau}) d_{\bk}\rangle$ is the 
average $d$-fermion number (spin) density, $g_n=\tilde{g}(1-\lambda)$, $g_s=\tilde{J}\lambda$, and 
we have used ${\cal S}_{\br}\cdot{\cal S}_{\br}=\sum_{\alpha\beta\gamma\delta}d^{\dagger}_{\br\alpha}
d_{\br\beta}d^{\dagger}_{\br\gamma}d_{\br\delta}(2\delta_{\alpha\delta}\delta_{\beta\gamma}-
\delta_{\alpha\beta}\delta_{\gamma\delta})$ to evaluate the mean-field contributions from the 
spin-dependent interaction. 
Although the Hartree and exchange contributions individually depend on $\lambda$, the Hartree-Fock 
mean-field Hamiltonian is {\it independent} of this arbitrary parameter, thereby satisfying a minimum 
requirement for physically meaningful conclusions. In contrast, a naive HS approach which includes 
only the Hartree (or exchange) self-energy gives an unphysical $\lambda$-dependent mean-field 
Hamiltonian~\cite{littlewood,bedell}. 

The resulting Hartree-Fock Hamiltonian  
\begin{eqnarray}
\label{eq: two}
H_{MF} & = & \sum_{\bk} d^{\dagger}_{\bk\sigma}\left[\xi_\bk\tau^{z}-\frac{\tilde{g} M}{2}\mathbf{1}-
\Delta\tau^{x}\right]_{\sigma\sigma'}d_{\bk\sigma'}
\end{eqnarray} 
is easily diagonalized to yield the quasiparticle energies 
$E_{\pm}(\bk)=-\tilde{g}M/2\pm\sqrt{\xi^2_\bk+\Delta^2}$. Here $M$ is the ferromagnetic 
order-parameter and $\Delta=\tilde{g}m_x/2$ is (purely real) superconducting order-parameter. The 
self-consistent equations for $M$ and $\Delta$ are 
\begin{eqnarray}
\label{eq: three}
M & = & \int \frac{d^3k}{(2\pi)^3}\left[1-n_{+}(\bk)-n_{-}(\bk)\right],\\
\label{eq: four}
1 & = & 2 \tilde{g} \int \frac{d^3k}{(2\pi)^3} \frac{n_{-}(\bk)-n_{+}(\bk)}
{\sqrt{\xi^2_\bk+\Delta^2}}.
\end{eqnarray}
These equations are similar to Eqs.(6) and (7) in~\cite{littlewood} {\it but contain only one 
effective coupling constant $\tilde{g}$}.  For $\tilde{g}<0$ Eq.(\ref{eq: four}) implies that 
$\Delta=0$, and it follows from Eq.(\ref{eq: three}) that $M \neq 0$ solutions can occur only if 
$\tilde{g}\leq\tilde{g}_c$ where $\tilde{g}_c$ is determined by Stoner's criterion. For $\tilde{g}>0$ 
we get the BCS solution, $\Delta\propto\exp\left(-1/\tilde{g}\cal{N}\right)$, and 
Eq.(\ref{eq: three}) implies that $M=0$. 

We conclude that coexistence of superconductivity and ferromagnetism requires physics beyond that of a 
single-band model with short-range interactions, and that HS based mean-field approximations must be 
used with caution~\cite{kerman,jm}, especially when separate terms in the interaction Hamiltonian are 
represented by different auxiliary fields.  




\begin{thebibliography}{99}
\bibitem{caveat} We do not concentrate on the energies or regimes-of-validity of various mean-field 
phases~\cite{zhou,shen} because the HS mean-field theory breaks down at an earlier stage.
\end{thebibliography}
\end{document}